\documentclass{desyproc}

\begin{document}
\title{ALP hints from cooling anomalies}

\author{{\slshape Maurizio Giannotti}\\[1ex]
Barry University, Miami Shores, US}

\contribID{giannotti\_maurizio}

\confID{11832}  
\desyproc{DESY-PROC-2015-02}
\acronym{Patras 2015} 
\doi  

\maketitle

\begin{abstract}
We review the current status of the anomalies in stellar cooling and argue that, among the new physics candidates, an axion-like particle would represent the best option to account for the hinted additional cooling.
\end{abstract}

\section{Introduction}

For over 2 decades, observations of different stellar systems have shown deviations from the expected behavior, indicating in all cases an over-efficient cooling. 

Statistically, each of these anomalies is not very significant. 
Taken together, however, they do seem to suggest the possibility of a common systematic problem in the modeling of the stellar evolution, in particular of the cooling mechanisms.

Is this a hint to physics beyond the standard model? 
If so, what kind of new physics?
As we shall see, among the various options the axion, or Axion-Like-Particle (ALP), solution is the most appealing and, in fact, the one most frequently considered in the past.

The axion~\cite{Weinberg:1977ma,Wilczek:1977pj} is a light pseudoscalar particle predicted by the most widely accepted solution of the strong CP problem~\cite{Peccei:1977hh,Peccei:1977ur} and a prominent dark matter candidate~\cite{Abbott:1982af,Dine:1982ah,Preskill:1982cy}.
Its interactions with photons and fermions are described by the Lagrangian terms
\begin{eqnarray}
 L_{\rm int}=- \frac{1}{4}g_{a\gamma} \, a F_{\mu \nu} \tilde F^{\mu \nu} - \sum_{\rm fermions}g_{ai} \, a \overline \psi_i \gamma_5 \psi_i \,,
\label{eq:Lint}
\end{eqnarray}
where $ g_{a\gamma}= C_\gamma \alpha/2\pi f_a$ and $ g_{ai}= C_i m_i/f_a$, with $ C_\gamma$ and $ C_i $ model dependent parameters and $ f_a $ a phenomenological scale known as the Peccei-Quinn symmetry breaking scale.

Moreover, in the so called QCD axion models, mass and interaction scale (Peccei-Quinn constant) are related as $ (m_{a}/1\mbox{ eV})= 6\times 10^6 {\rm GeV}/f_a $. 
This describes a band (the width given by the possible values of the model dependent parameters) in the mass-coupling (e.g., to photons) parameter space, known as the QCD axion line.
Belonging to this band, however, is not a requirement for the solution of the strong CP problem~\cite{Rubakov:1997vp,Berezhiani:2000gh,Gianfagna:2004je}.

More general models of pseudoscalar particles, known as ALPs, 
which couple to photons (and, possibly, to fermions) but do not satisfy the above mass-coupling relation, emerge naturally in various extensions of the Standard Model though, in general, their existence is not related to the strong CP problem~\cite{Ringwald:2012hr}. 

If appropriately coupled to electrons, photons, and nucleons, ALPs could explain the stellar cooling anomalies. 
Additionally, light ALPs have been invoked to explain other astrophysical anomalies, such as the seeming transparency of the universe to very high energy (TeV) gamma rays in the galactic and extragalactic medium~\cite{Horns:2012fx} and some anomalous redshift-dependence of AGN gamma-ray spectra~\cite{Galanti:2015rda} (though this last hypothesis currently shows some conflict with the SN bound on the axion-photon coupling~\cite{Payez:2014xsa}).
More recently, it was also pointed out that anomalous X-ray observations of the active Sun suggest an ALP-photon coupling~\cite{Rusov:2015sqa} of the same size hinted by the other analyses.

Interestingly, the required couplings are not excluded by experiments nor by phenomenological considerations and are accessible to the new generation ALP detectors, in particular ALPS II~\cite{Bahre:2013ywa} and the International Axion Observatory (IAXO)~\cite{Irastorza:2011gs,Vogel:2013bta}.

\section{Observational anomalies is stellar cooling and ALPs}

\subsection{White dwarfs}

\begin{wrapfigure}{R}{0.5\textwidth}
   \centerline{\includegraphics[width=0.45\textwidth]{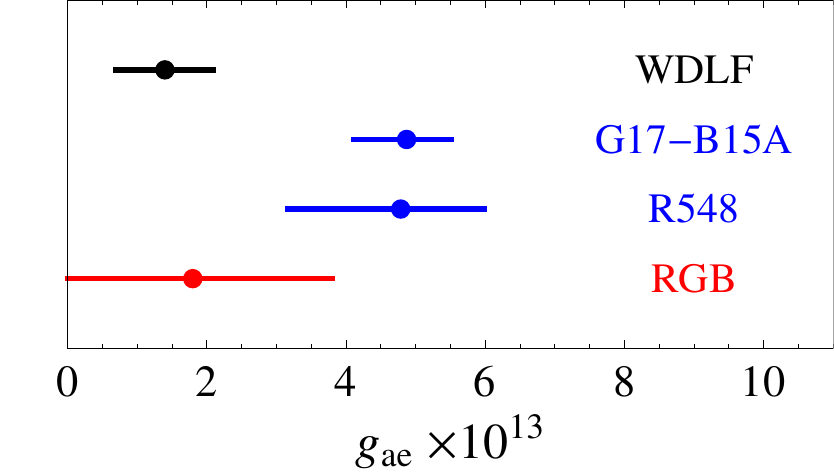}}
\caption{Summary of hints on the ALP-electron coupling from WD and RGB stars (at $ 1\sigma $).
}
\label{fig:gae}
\end{wrapfigure}

For over two decades, observations of the period decrease ($ \dot P /P $) of particular white dwarf (WD) variables have shown discrepancies (at $ 1\sigma $) with the expected behavior.
In particular, all the variables studied (two pulsating DA WDs, G117-B15A~\cite{Corsico:2012ki,BischoffKim:2007ve} and R548~\cite{Corsico:2012sh}, and one pulsating DB WD, PG 1351+489~\cite{Corsico:2014mpa}) show an unexpectedly fast cooling ($ \dot P /P $ is practically proportional to the cooling rate $ \dot T /T $), suggesting the possibility of additional energy loss channels.
The results from the two DA WD show a preference for an axion coupled to electrons with 
$ g_{ae}\simeq 4.8\times 10^{-13} $~\cite{Corsico:2012ki,Corsico:2012sh} (see Fig.~\ref{fig:gae}).
The no-axion solution is recovered at 2$ \sigma $.

Additionally, various studies of the WD luminosity function (WDLF), which represents the WD number density per brightness interval, also seem to indicate a preference for an additional cooling channel and, in particular, for an axion-electron coupling 
$ g_{ae}\simeq (1.4\pm 0.3) \times 10^{-13} $ (at 1$ \sigma $)~\cite{Bertolami:2014wua}.
A more recent study of the hot part of the WDLF~\cite{Hansen:2015lqa} did not confirm this anomalous behavior.
However, the hotter section of the WDLF has much larger observational errors and the ALP production would be almost completely hidden by standard neutrino cooling in the hottest WDs. 

It should also be  noted that the hints on the axion-electron coupling from the WDLF and the WD pulsation disagree at $ 1\sigma $ indicating, perhaps, an underestimate of the errors. 
In particular, the results from the pulsating WDs are based on assumptions on the analyzed oscillating mode that should be independently verified (see, e.g., discussion in~\cite{Bertolami:2014wua}).

\subsection{Red giants}

Further hints to anomalous energy loss in stars emerge from the recent analysis of the Red Giant Branch (RGB) stars in~\cite{Viaux:2013hca,Viaux:2013lha}.
This showed a brighter than expected tip of the RG branch in the M5 globular cluster, indicating a somewhat over-efficient cooling during the evolutionary phase preceding the helium flash.

The anomalous brightness, $ \Delta M_{I,{\rm tip}}\simeq 0.2 $ mag in absolute $ I- $band magnitude, is within the calculated observational and modeling errors, which include uncertainties in the mass loss, treatment of convection, equation of state and cluster distance.
However, the error budget seems to just barely compensate for the difference between observed and expected brightness.
A better agreement would require an anomalous cooling of a few $10^{33}$ erg/s, which could be accounted for by a neutrino magnetic moment 
$ \mu_\nu\sim (1 - 2)\times 10^{-12} \mu_{\rm B} $~\cite{Viaux:2013hca}, where $ \mu_{\rm B} $ is the Bohr magneton,
or an axion-electron coupling  
$ g_{ae} \sim (1 - 2) \times 10^{-13}$~\cite{Viaux:2013lha}.

A reduction of the uncertainties,  particularly a better determination of the cluster distance, which may become possible with the GAIA mission, will certainly help clarifying the physical significance of this discrepancy.

\subsection{Horizontal branch stars}

A recent analysis~\cite{Ayala:2014pea} showed a mild disagreement (at 1$ \sigma $) between the observed and the expected R-parameter, 
$R= {N_{\rm HB}}/{N_{\rm RGB}}$, 
which compares the numbers of stars in the HB  ($N_{\rm HB}$) and in the upper portion of the RGB
($N_{\rm RGB}$).
More specifically, the observed value, $ R=1.39\pm 0.03 $ is somewhat smaller than the expected one 
$ 1.44\leq R\leq 1.50$. 

The higher than expected value of $ R $ indicates a surplus of HB stars with respect to RGB in the examined clusters,
suggesting that HB stars are cooling more efficiently, and therefore are less numerous, than expected. 

This result may be due to an ALP coupled to photons with $ g_{a\gamma}=(0.29-0.57)\times 10^{-10} {\rm GeV}^{-1}$.
A more recent analysis (see O. Straniero's contributions to these proceedings) indicates a slightly smaller value for the hinted coupling but preserves the discrepancy at the $ 1\sigma $ level.

\subsection{Massive he-burning stars}

Another long standing puzzle is the smaller than predicted number ratio of blue over red supergiants in open clusters (see~\cite{McQuinn:2011bb} and references therein).

Stars of mass a few times larger than the sun, during their helium burning stage evolve from red (cold) to blue (hot) and back. 
This journey is called the blue loop and is very sensitive to the microphysics governing the stellar evolution, in particular its cooling mechanism. 
The observation of less blue stars indicates a shorter than expected blue stage, which can be attributed to a more efficient than expected cooling of the core~\cite{Friedland:2012hj,Carosi:2013rla,Giannotti:2014cpa}.

The analysis in~\cite{Friedland:2012hj} indicated that an axion-photon coupling of a few $ 10^{11} {\rm GeV}^{-1}$, in the same range as the one hinted by the HB anomaly, would reduce the number of expected blue stars, alleviating or perhaps solving the anomaly. 
However, in this case the uncertainties in the microphysics and in the observations are, essentially, unquantifiable.

\subsection{Neutron stars}

Finally, x-rays observations of the surface temperature of a neutron star in Cassiopeia A also showed a cooling rate considerably faster than expected.
The effect may be interpreted in terms of an axion-nucleon coupling of the order of
$ g_{an}\sim 4\times 10^{-10} $~\cite{Leinson:2014ioa}.

However, the uncertainties in the physics of neutron stars cooling make this only a marginal hint.
Indeed, the effect could have a different origin, for example as a phase transition of the neutron condensate into a multicomponent state~\cite{Leinson:2014cja}.

\section{Is this an ALP?}

Among the \textit{new physics} explanations, the existence of ALPs is the most appealing and the most frequently invoked.
To explain the cooling anomalies, ALPs should couple to photons and fermions, as in Eq.~(\ref{eq:Lint}) with, for example, $ f_a\simeq 10^7 $GeV,  $ C_\gamma\sim 1 $ and $ C_{e}\sim C_{n}\sim 10^{-2}$~\cite{Ringwald:2015lqa}.

A study (in preparation) shows that none of the other common candidates can explain the combined observed deviations from the standard cooling of the diverse stellar systems.
In particular, an anomalous neutrino magnetic moment has essentially no effects on the WDLF~\cite{Bertolami:2014noa}.
Moreover, even if equipped with a magnetic moment as large as currently allowed
by experimental limits and astrophysical observations, 
neutrinos would not be effectively produced in low density stars, such as HB or massive He burning stars.

Analogously, preliminary results show that the regions of the hidden photon (HP) parameter space necessary to explain the HB and RGB anomalies do not overlap and the region in which HP could reconcile the WDLF observations is phenomenologically excluded.

\section{Summary and conclusion}

Numerous independent observations seem to indicate excessive energy loss in several stellar system.
The combination of the anomalous observations of WD, RG and HB stars, strongly favors  ALPs with respect to other possible candidates.

Additionally, ALPs have been invoked for the solution of other unexplained astrophysical observations. 
Most importantly, the quest for dark  matter, of which the axion provides an excellent candidate (see P. Sikivie's contribution to these proceedings). 
Additionally, a light ALP coupled to photons has been proposed to explain observations of the seeming transparency of the universe to very high-energy gamma-rays and an anomalous redshift-dependence of AGN gamma-ray spectra.

Remarkably, the most important section of the hinted ALP parameter space could be investigated with the next generation of axion detectors. 
%
A discovery of an ALP in the parameter region discussed would be revolutionary not only for particle physics and probably for cosmology, but also for TeV gamma ray astronomy and for stellar evolution.

%

\section{Bibliography}


\begin{footnotesize}

\end{footnotesize}


\end{document}